# The Brightness of VisorSat-Design Starlink Satellites

Anthony Mallama

anthony.mallama@gmail.com

Abstract

The mean of 430 visual magnitudes of VisorSats adjusted to a distance of 550-km (the operational altitude) is 5.92 +/-0.04. This is the characteristic brightness of these satellites when observed at zenith. VisorSats average 1.29 magnitudes fainter than the original-design Starlink satellites and, thus, they are 31% as bright.

## 1. Introduction

The rapidly growing population of bright satellites in low-earth-orbit is beginning to interfere with ground-based astronomy (Otarola et al., 2020, Walker et al., 2020, Tyson et al., 2020, Gallozzi et al., 2020, Hainaut and Williams, 2020 and McDowell, 2020). The SpaceX company has launched hundreds of Starlink communication satellites during the past two years and plans to put thousands more into orbit in the near future. OneWeb is actively pursuing a similar project and other companies may follow.

SpaceX has engaged with the astronomical community to address this issue. As a result they have developed a VisorSat model of Starlink that includes a Sun shade to make the satellites appear dimmer. This paper characterizes the observed brightness of VisorSat-design spacecraft in visible light and compares it to that of the original Starlink design.

Section 2 of this paper defines the magnitudes used herein and discusses the illumination phase angle which can affect brightness. Then Section 3 describes the physical shape of Starlink satellites and discusses their orbits as these factors pertain to brightness. Section 4 describes the sources of satellite magnitudes used in this study as well as the data processing techniques. Section 5 presents the findings which include the mean VisorSat magnitude, its uncertainty, the dispersion and the phase function. Section 6 compares the brightness of VisorSat with that of the original-design Starlink satellites and with the OneWeb satellite constellation. Section 7 discusses some limitations of this study. Section 8 briefly addresses VisorSats below the operational altitude as well as brightness surges. Section 9 gives the conclusions.

This study is part of the author's work with the Satellite Observations Working Group (Otarola et al., 2020). Most of the SOWG members record observations made through spectral filters. The visual magnitude results in this paper characterize satellite brightness across all visible wavelengths and complement the color-filtered photometry being carried out by other members of the SOWG.

## 2. Magnitudes and phase angles

*Visual* magnitudes represent satellite brightness in combined blue, green and red light with an effective wavelength in green. The band-pass sensitivity is centered near 0.6 micron and its width is about 0.4 micron. Visual magnitudes are a close match to the V-band of the Johnson-Cousins photometric system although the bandwidth of V is narrower.

The *apparent magnitude* is that recorded by a sensor after correction for atmospheric extinction. An analogy to the absolute magnitude of astronomers is the *1000-km magnitude* used by the satellite community. The 1000-km magnitude is an adjustment of apparent brightness to that standard distance which allows for comparisons between multiple satellite constellations at different altitudes.

Finally, there is a *standard magnitude* that is intended to account for the illumination phase angle which is measured at the satellite between the Sun and the observer. In principle the standard magnitude gives the brightness of a half-illuminated satellite at 1000 km distance. The adjustment between the 1000-km magnitude and the standard magnitude is given in Equation 1,

$$C = 2.5 * \log10 ( 0.5 + 0.5 \cos B ) + 0.753$$

Equation 1

where $C$ is the correction to be added. For an observer at phase angle $B$ the term inside the parentheses represents the fraction of a spherical object that is illuminated by the Sun. The value 0.753 sets the correction to zero for a satellite at $B = 90^{O}$. In practice the standard magnitude is of limited value for Starlink satellites because they are not shaped like spheres.

A potentially useful approach for Starlink is to derive an empirical phase function by determining a least-squares fit between the 1000-km magnitudes and their phase angles. An empirical fit for VisorSat magnitudes is presented later in this paper.

The SOWG generally employs the Minnaert (1941) bidirectional reflectance function to process observed magnitudes. Other formalisms can be used, too. Cole (2020) developed a physical model for VisorSats that includes its solar array.

## 3. Starlink satellites and their orbits

The orbital characteristics of Starlink satellites along with their physical size, shape, attitude (that is, their orientation in space) and reflective properties determine their brightness. Spacing between the observer and the satellite affects brightness according to the inverse square of the distance. SpaceX plans to distribute Starlink satellites into several orbital 'shells' of different altitudes and, thus, different distances. To date all satellites have been assigned to the 550 km altitude shell.

Starlink satellites are launched in batches of about 60 at a time on a single rocket. They are initially injected into lower altitude orbits where they reside until precession advances them to the desired celestial plane. Then they ascend to their final orbital altitude in groups of about 20 where they are spaced apart by 5 minutes of time and orbit the Earth in 95 minutes. These groups of 20 may be observed from one location on a single night.

The satellites themselves consist of a flat-panel shaped bus and a large attached solar array. The flat panel is nominally oriented perpendicular to the nadir direction and the solar panel is positioned to intercept sunlight. The bus appears to have both diffusive and specularly reflective surfaces.

Starlink-1436 was the first VisorSat-design satellite to be launched. In this model a Sun shade was added to the bus in order to reduce the amount of sunlight reflected from the nadir side toward observers on the Earth's surface. That initial VisorSat was orbited on 2020 June 4 on the seventh operational Starlink launch designated L7. Beginning with the L9 launch of 2020 August 7 all Starlink satellites have been VisorSats.

## 4. Observation and data processing

The magnitudes analyzed in this paper are visual as defined in Section 2. That is, they represent the brightness of Starlink satellites in combined red, green and blue colors with an effective wavelength near 0.6 micron.

The data come from three sources. Two of them are experienced visual observers using binoculars. J. Respler records satellite magnitudes from coordinates $40^{O}$ north and $74^{O}$ west while the author observes from $39^{O}$ north and $77^{O}$ west. Satellite brightness is determined by comparison with nearby stars of known magnitudes using interpolation. The observers aim to achieve magnitudes that are accurate to about 0.2 magnitudes. Each apparent magnitude is converted to a 1000-km magnitude based on the satellite range at the time of observation. An associated phase angle is determined from the positions of the satellite, Sun and observer.

On 7 occasions out of 290 binocular observations the satellite was too faint to be seen. Since the limiting magnitude is usually about 8 and few satellites are close to that limit, the value 8.5 was assigned for those 7 observations. Figure 1 illustrates the distribution of apparent magnitudes.

The third source of data is the Mini-MegaTORTORA (MMT) automated observatory (Karpov et al. 2015) in Russia at $44^{O}$ north and $41^{O}$ east. Each of the 9 MMT channels consists of a 71 mm diameter f/1.2 lens and a 2160 x 2560 sCMOS detector. Unfiltered MMT magnitudes are within 0.1 magnitude of the Johnson-Cousins V band-pass for objects with small B-V color indices according to a color transformation formula kindly provided by S. Karpov (private communication). Mallama (2020a) found MMT magnitudes to be consistent with those of visual observers.

The online MMT database lists standard magnitudes which include the phase angle correction for a spherical body described in Section 2. Since Starlink satellites are not spherical the correction indicated by Equation 1 is removed in order to give the 1000-km magnitude for the purpose of this study. When a satellite is recorded simultaneously by two separate MMT channels the magnitudes may differ by 0.1 - 0.2 magnitude which suggests that accuracy is of that order.

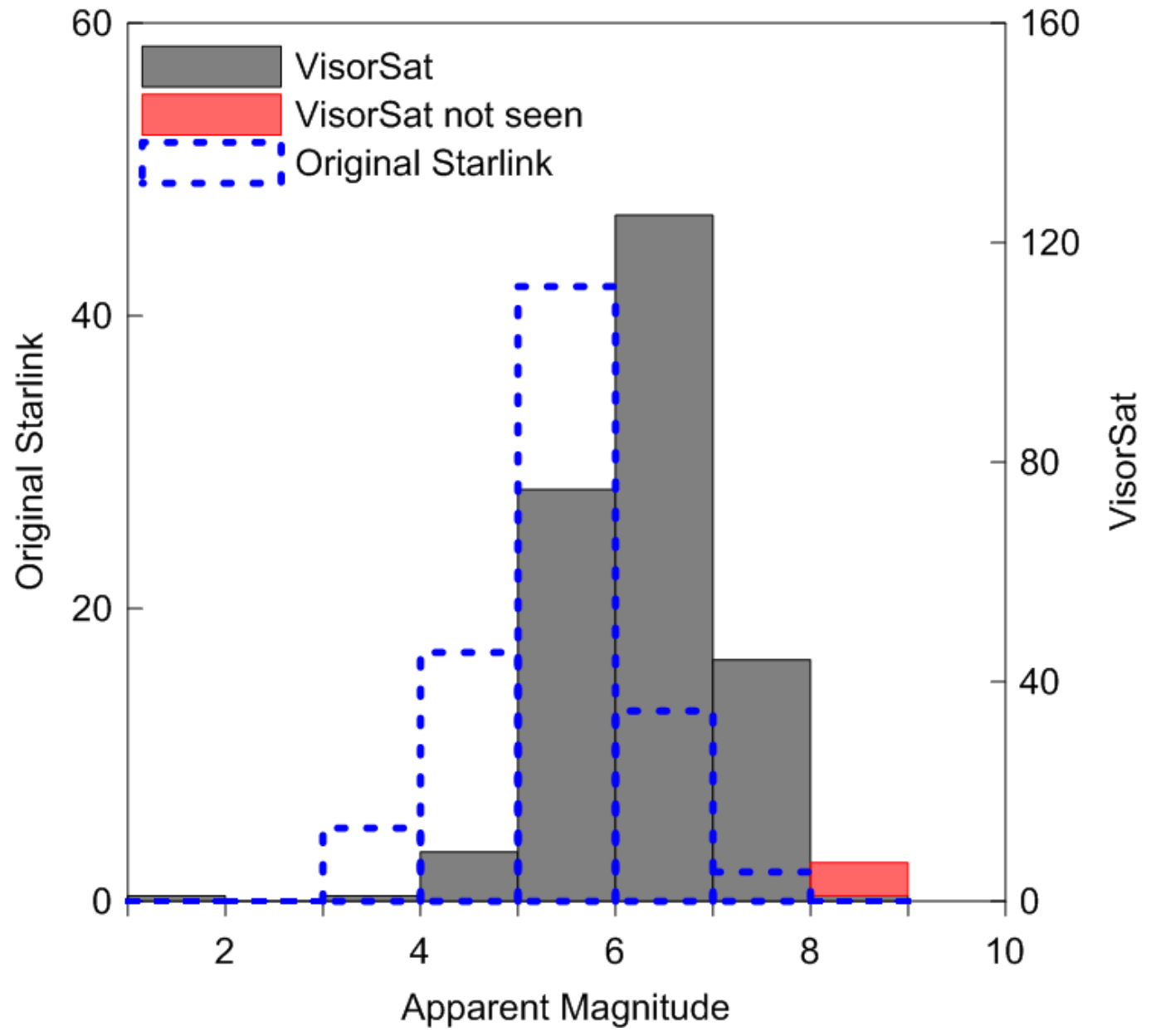

Figure 1. Histogram of apparent VisorSat magnitudes recorded by the visual observers. Also shown are the assigned magnitudes where the satellites were too faint to be seen, along with apparent magnitudes for original-design Starlink satellites.

Starlink satellites spend the great majority of their lifetime in orbit at the 550 km operational altitude. Furthermore, SpaceX (2020) indicates that they adjust the configuration of the solar panel differently during the ascent and the operational phases. Therefore, only observations of satellites at the operational altitude were selected for the analysis in the next section, while observations of satellites at lower altitudes are addressed later. Measurements taken when the satellite was in the Earth's penumbral shadow were not used. Each of the 430 observations used in Section 5 sampled a different satellite pass. The data are listed in the Appendix A.

5. Brightness characterization

The mean 1000-km magnitude for satellites at the 550 km operational altitude is 7.22 +/- 0.04 and the standard deviation of the individual magnitudes is 0.85. The median magnitude is somewhat fainter at 7.36 indicating that the mean is skewed by a few very bright observations. The mean adjusted to a range of 550 km gives magnitude 5.92 which is an indication of VisorSat brightness at zenith.

Figure 2 shows the magnitudes plotted as a function of phase angle. The linear and quadratic fits indicate that the satellites are brighter at small phase angles as expected. However, the standard deviation remains at 0.085 after the linear fit is applied and it only reduces to 0.084 for the quadratic fit. So phase functions are of little practical value in predicting satellite brightness. The best fitting equations are given in Table 1.

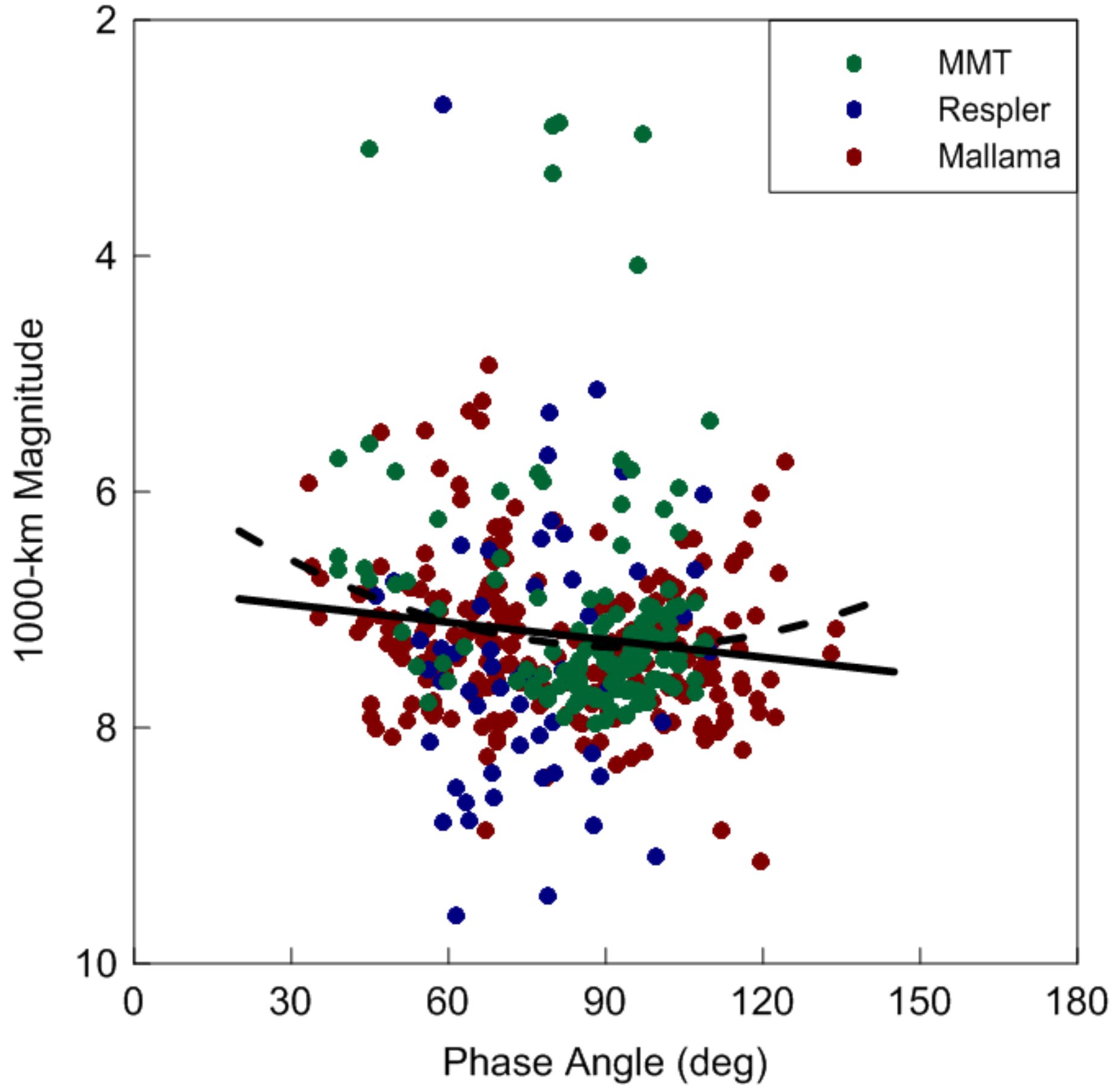

Figure 2. The phase function with linear and quadratic least-squares fits.

Observations of original-design Starlink satellites also evidenced large observed scatter with little reduction from modeling. Otarola et al. (2020) determined a standard deviations of 0.75 magnitude for POMENIS V-band photometry of satellites above 550 km before normalization techniques were applied and that only reduced to 0.69 afterward. Likewise, Mallama (2020a) derived 0.67 for 1000-km original-design Starlink magnitudes and 0.65 after a linear phase angle correction was applied.

Table 1. Phase Functions

| | |
|---|---|
| Linear | $M_V = 6.809 + 0.00495 * P$ |
| Quadratic | $M_V = 5.732 + 0.03352 * P - 0.000177 * P^2$ |

$M_V$ = visual magnitude

P = phase angle in degrees

The large dispersion values are expected. Visual observers have often noticed that Starlink satellites viewed in succession with nearly identical satellite-Sun-observer geometries can vary widely in brightness. For example, J. Respler wrote "there were 4 pairs [of satellites] with second following first by several seconds. In each case the first was mag 4-6. The second was 2-3 mags brighter" (http://www.satobs.org/seesat/Mar-2020/0126.html).

Additionally, MMT light curves demonstrate substantial and seemingly erratic magnitude changes as illustrated in Figure 3. The large brightness variations seen by visual observers and recorded by MMT are probably due to satellite attitudes. Thus, accurate prediction of the brightness of any given satellite at any given time may remain a difficult problem to solve.

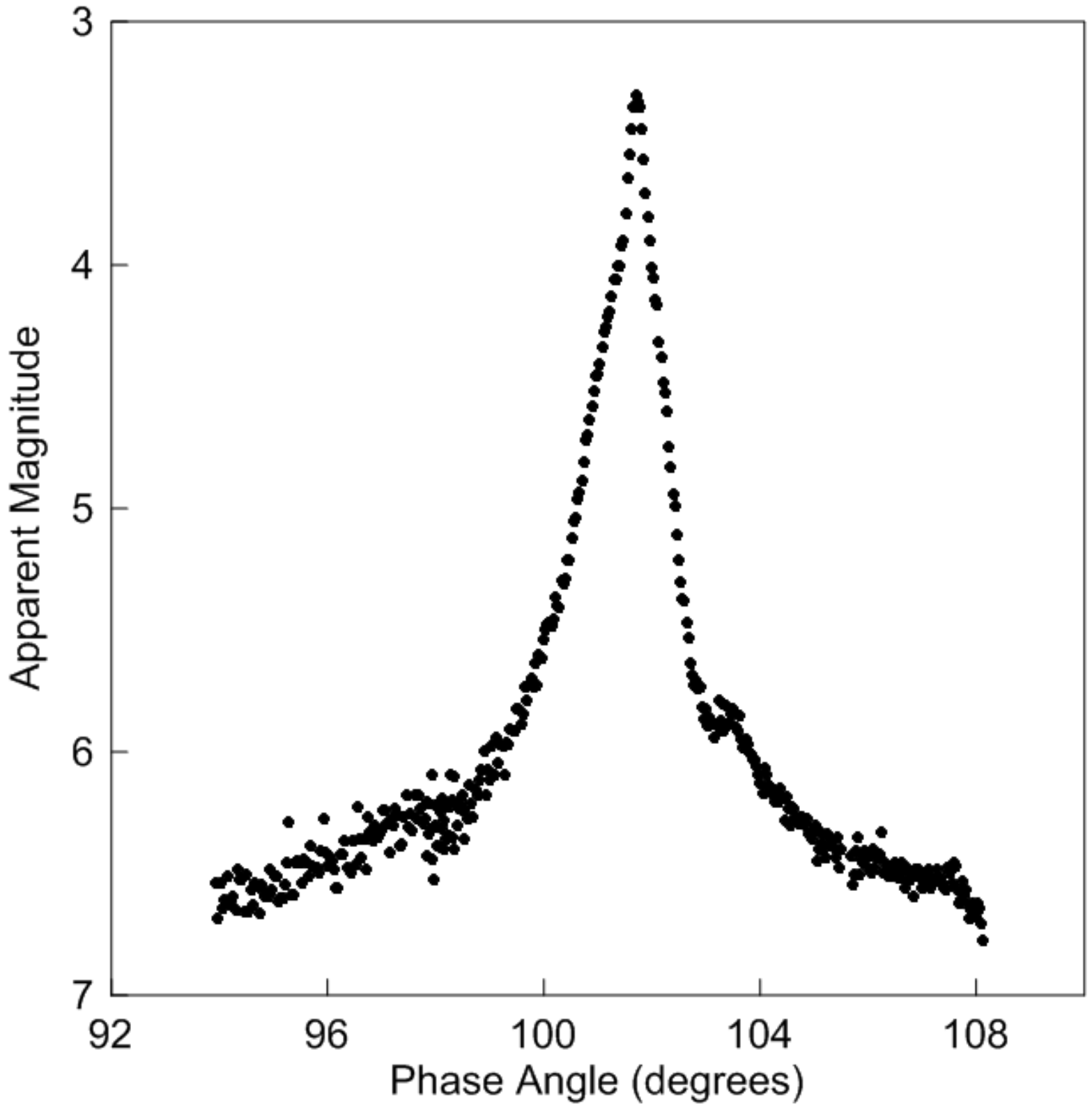

Figure 3. MMT light curve of Starlink-1549 recorded on 2020 November 27. The large brightness surge is not explained by the phase angles of the observations.

Two other independent variables besides phase angle have been examined. Figure 4 shows magnitudes plotted versus calendar days. Meanwhile, Figure 5 shows magnitudes versus days elapsed since a satellite reached the 550 km operational altitude. Both of these functions seem to indicate dimming over time and these effects may be aliasing. The best fit to the calendar day plot indicates a fading of 0.52 magnitude between day 236 and 363 of calendar year 2020. The power law fitted to the magnitudes in Figure 5 suggests about half a magnitude of fading shortly after the satellites attain 550 km altitude.

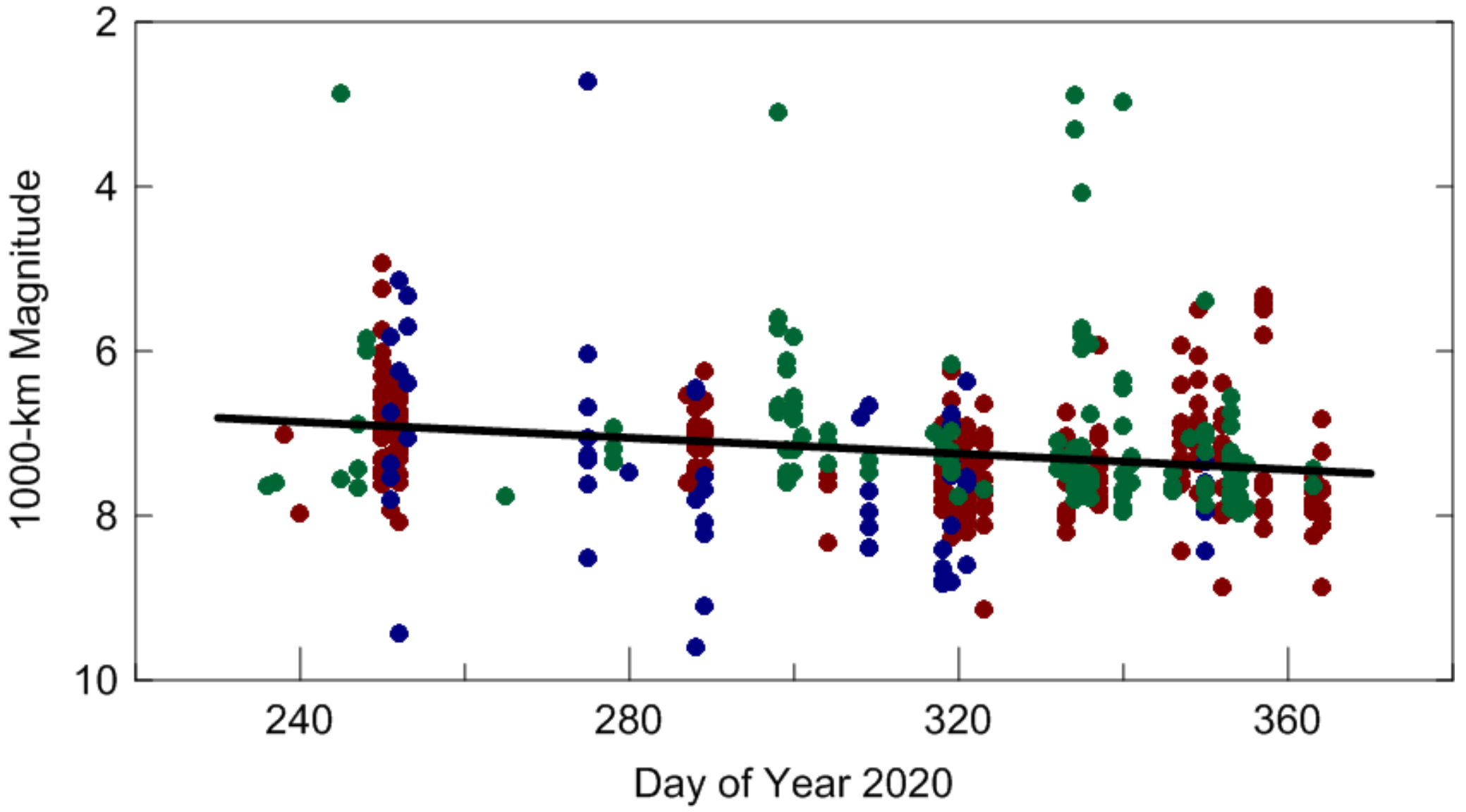

Figure 4. Brightness as a function of time.

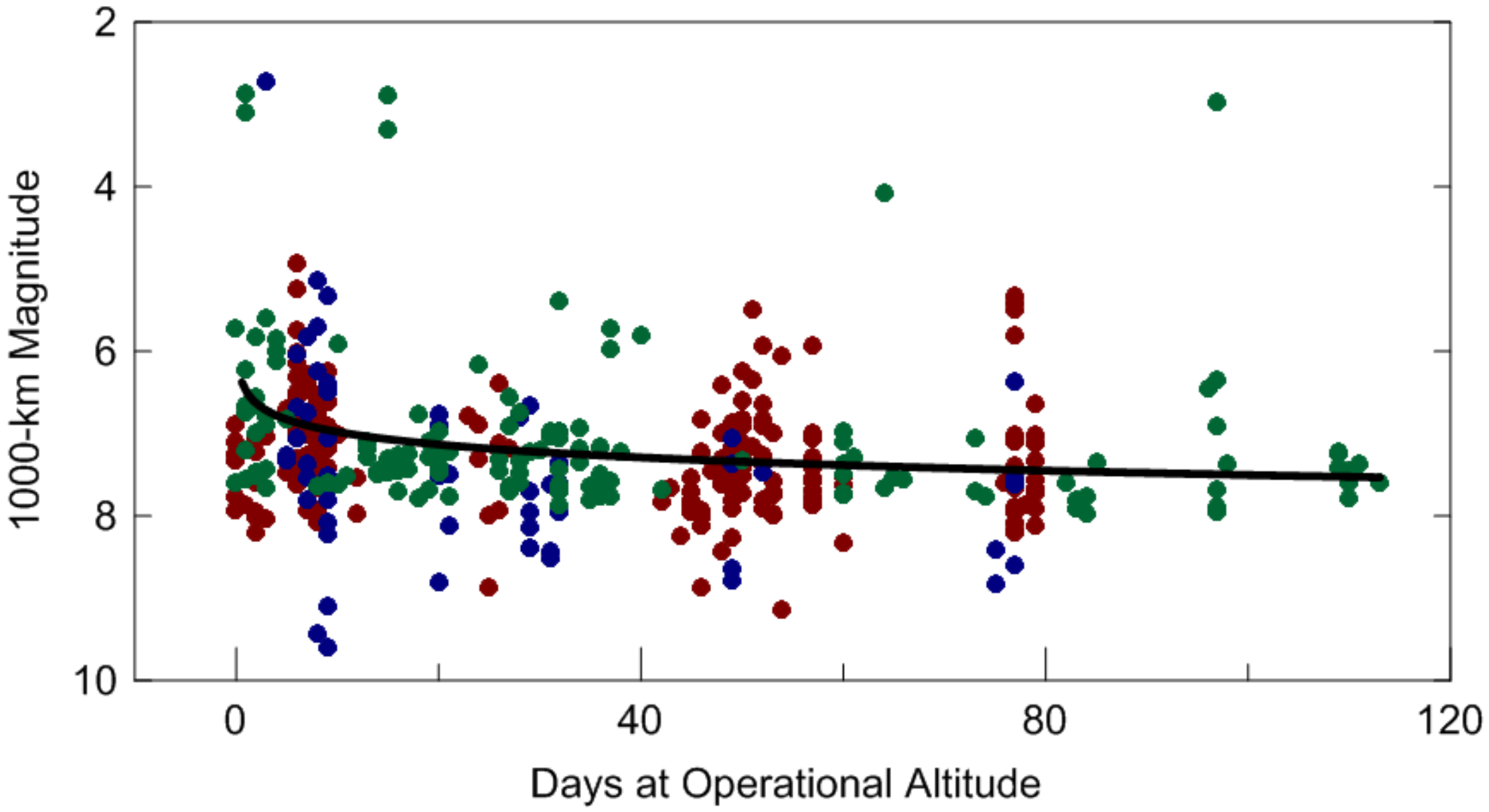

Figure 5. Brightness versus time after reaching 550 km altitude fitted with a power law.

## 6. Comparison with original-design Starlink and with OneWeb

There are several hundred original-design Starlink satellites in orbit. The mean VisorSat magnitude reported in the previous section is 1.29 fainter than that of the original-design satellites as reported by Mallama (2020a). Thus, VisorSats average 31% as bright as their predecessors.

OneWeb satellites are another constellation being launched in great numbers. These spacecraft have large solar arrays on long support arms and dish-like antennas on shorter arms. Their operational altitude of 1,200 km is more than twice that of Starlink. Mallama (2020b) determined a mean 1000-km magnitude of 7.18 +/-0.03 for OneWeb based on MMT observations. An analysis of visual binocular observations which were reported at a later time (satobs.org/seesat/Dec-2020/0072.html) gave results that are consistent with those from MMT. The 1000-km value adjusted to the nominal 1,200 km altitude of a OneWeb satellite in orbit corresponds to magnitude 7.58. Thus, VisorSats satellites are approximately the same brightness as OneWeb at a common distance but they are much brighter than OneWeb at their respective operational altitudes. Table 2 summarizes these comparisons.

Table 2. Magnitude comparisons

| Satellite | 1000-km | Operational Altitude | Uncertainty |
|---|---|---|---|
| VisorSat | 7.22 | 5.92 | +/-0.04 |
| Original Starlink | 5.93 | 4.63 | +/-0.02 |
| OneWeb | 7.18 | 7.58 | +/-0.03 |

## 7. Limitations of this study

Several factors place limits on the accuracy and the general applicability of the results in this paper. First is the geographic distribution of the three data sources noted in Section 3. The MMT is located in Russia and the two visual observers are located in the eastern United States. If spacecraft attitude is being adjusted geographically for operational reasons then the results presented here may not be representative of other regions. Furthermore, the Sun's varying declination throughout the year may affect satellite brightness in a complicated way that depends on a combination of season and geographic location.

Another limiting factor is the short time span of the observations which begins in 2020 August and ends in December. Since spacecraft attitude can be actively controlled, the observed magnitudes during this period may not reflect those at later times.

Lastly, the MMT standard magnitudes and phase angles used in the analysis are averages of those for each satellite track. The magnitudes were copied from the MMT database while the phase angles were estimated by eye from track plots. The estimated phase angles are accurate to about 1 degree. A more rigorous (and time consuming) approach would be to associate every magnitude in every track with its corresponding phase angle.

## 8. Lower altitude observations and flaring

Besides the data already discussed the author measured 22 magnitudes of VisorSats that were below the operational altitude. The mean 1000-km magnitude was 7.54 +/-0.03 and the mean altitude was 417 km. Thus, the satellites were not especially bright in an absolute sense although their mean apparent brightness was rather high at magnitude 5.65. The observations are listed in Appendix B.

The original Starlink satellites were observed to flare in brightness by up to 10 magnitudes on several occasions in 2020 April (Mallama 2020a). These reports were posted on the SeeSat-L mailing list at http://www.satobs.org. However, there have been no further reports of extreme flaring events for original or VisorSat Starlink satellites since that time.

## 9. Conclusions

Magnitudes of VisorSat satellites at their 550 km altitude measured by the MMT automated observatory in Russia and by two visual observers in the United States are analyzed. The main conclusion is that the mean of 430 visual magnitudes adjusted to a 550 km range is 5.92 +/-0.04. This is their characteristic brightness when seen at zenith. VisorSats average 1.29 magnitudes fainter than the original-design Starlink satellites. Thus, VisorSats are 31% as bright as their predecessors.

The following findings are also discussed. There is some indication of a modest brightness decline after the first few days of a satellite at 550 km and there may have been a general dimming over time in year 2020. Satellites below the operational altitude are not especially bright in an absolute sense but their apparent brightness is high. No extremely bright flaring of VisorSats has been reported since 2020 April.

Appendix A.

Summary of 550 km magnitudes
'Cal. Day' is the day number of year 2020.
'Ops. Days' is the time since attaining the 550km operational altitude.
Red entries indicate 'not seen' where apparent magnitude is assigned as 8.5

| Starlink Sat # | Cal. Day | Ops. Days | Phase Angle | 1000-km Magnitude | Source |
|---|---|---|---|---|---|
| 1436 | 238 | 10 | 71 | 7.02 | Mallama |
| 1436 | 240 | 12 | 86 | 7.96 | " |
| 1581 | 250 | 5 | 75 | 7.47 | " |
| 1526 | 250 | 6 | 75 | 7.52 | " |
| 1584 | 250 | 6 | 74 | 7.61 | " |
| 1523 | 250 | 6 | 73 | 7.01 | " |
| 1576 | 250 | 6 | 73 | 6.13 | " |
| 1580 | 250 | 6 | 72 | 7.45 | " |
| 1591 | 250 | 6 | 71 | 6.56 | " |
| 1534 | 250 | 6 | 69 | 6.79 | " |
| 1544 | 250 | 6 | 69 | 6.31 | " |
| 1565 | 250 | 6 | 68 | 4.93 | " |
| 1560 | 250 | 6 | 67 | 5.23 | " |
| 1556 | 250 | 6 | 94 | 6.95 | " |
| 1557 | 250 | 6 | 100 | 7.06 | " |
| 1567 | 250 | 6 | 94 | 7.31 | " |
| 1558 | 250 | 6 | 104 | 7.31 | " |
| 1569 | 250 | 6 | 106 | 7.42 | " |
| 1555 | 250 | 6 | 116 | 6.49 | " |
| 1582 | 250 | 6 | 120 | 6.01 | " |
| 1522 | 250 | 6 | 124 | 5.75 | " |
| 1581 | 250 | 5 | 123 | 6.69 | " |
| 1584 | 251 | 7 | 70 | 6.29 | " |
| 1523 | 251 | 7 | 70 | 6.39 | " |
| 1576 | 251 | 7 | 70 | 7.02 | " |
| 1580 | 251 | 7 | 69 | 7.93 | " |
| 1591 | 251 | 7 | 68 | 6.45 | " |
| 1534 | 251 | 7 | 67 | 6.85 | " |
| 1544 | 251 | 7 | 67 | 7.26 | " |
| 1565 | 251 | 7 | 67 | 7.44 | " |
| 1560 | 251 | 7 | 68 | 6.78 | " |
| 1556 | 251 | 7 | 77 | 6.75 | " |
| 1557 | 251 | 7 | 91 | 7.27 | " |
| 1567 | 251 | 7 | 95 | 7.45 | " |
| 1522 | 252 | 8 | 69 | 7.59 | " |
| 1581 | 252 | 7 | 69 | 6.48 | " |
| 1526 | 252 | 8 | 68 | 6.55 | " |
| 1584 | 252 | 8 | 68 | 6.86 | " |
| 1523 | 252 | 8 | 68 | 7.17 | " |
| 1576 | 252 | 8 | 69 | 6.59 | " |
| 1580 | 252 | 8 | 69 | 8.07 | " |
| 1591 | 252 | 8 | 69 | 6.95 | " |
| 1534 | 252 | 8 | 70 | 7.09 | " |
| 1544 | 252 | 8 | 72 | 7.30 | " |
| 1565 | 252 | 8 | 73 | 7.17 | " |
| 1560 | 252 | 8 | 75 | 7.50 | " |
| 1556 | 252 | 8 | 102 | 6.78 | " |
| 1557 | 252 | 8 | 100 | 6.72 | " |
| 1567 | 252 | 8 | 109 | 7.21 | " |
| 1611 | 287 | 8 | 56 | 7.59 | " |
| 1641 | 287 | 8 | 56 | 6.53 | " |
| 1597 | 288 | 8 | 55 | 7.45 | " |
| 1615 | 288 | 8 | 55 | 7.18 | " |
| 1629 | 288 | 8 | 56 | 6.69 | " |
| 1608 | 288 | 8 | 57 | 7.40 | " |
| 1628 | 288 | 8 | 69 | 7.65 | " |
| 1592 | 288 | 8 | 81 | 7.40 | " |
| 1626 | 288 | 8 | 93 | 6.93 | " |
| 1601 | 288 | 8 | 96 | 7.43 | " |
| 1630 | 288 | 8 | 103 | 6.95 | " |
| 1595 | 288 | 9 | 56 | 7.16 | " |
| 1640 | 288 | 9 | 56 | 7.58 | " |
| 1611 | 288 | 9 | 57 | 6.91 | " |
| 1641 | 288 | 9 | 58 | 7.01 | " |
| 1597 | 288 | 9 | 59 | 6.90 | " |
| 1615 | 289 | 9 | 62 | 7.41 | " |
| 1629 | 289 | 9 | 102 | 7.16 | " |
| 1608 | 289 | 9 | 95 | 7.17 | " |
| 1628 | 289 | 9 | 100 | 6.93 | " |
| 1592 | 289 | 9 | 90 | 6.95 | " |
| 1626 | 289 | 9 | 92 | 6.99 | " |
| 1601 | 289 | 9 | 115 | 6.59 | " |
| 1630 | 289 | 9 | 118 | 6.23 | " |
| 1613 | 289 | 9 | 114 | 6.61 | " |
| 1567 | 304 | 60 | 92 | 8.32 | " |
| 1558 | 304 | 60 | 94 | 7.51 | " |
| 1569 | 304 | 60 | 96 | 7.61 | " |
| 1542 | 318 | 0 | 92 | 7.93 | " |
| 1551 | 318 | 0 | 108 | 6.88 | " |
| 1568 | 318 | 0 | 111 | 7.25 | " |
| 1559 | 318 | 0 | 114 | 7.09 | " |
| 1542 | 319 | 1 | 113 | 7.86 | " |
| 1578 | 319 | 0 | 116 | 7.32 | " |
| 1536 | 319 | 0 | 119 | 7.76 | " |
| 1530 | 318 | 49 | 72 | 7.05 | " |
| 1527 | 318 | 49 | 71 | 7.46 | " |
| 1624 | 318 | 49 | 48 | 7.28 | " |
| 1577 | 318 | 49 | 51 | 7.41 | " |
| 1574 | 318 | 49 | 53 | 7.80 | " |
| 1554 | 318 | 49 | 57 | 7.90 | " |
| 1541 | 318 | 49 | 79 | 7.69 | " |
| 1543 | 318 | 49 | 99 | 7.59 | " |
| 1535 | 318 | 49 | 102 | 7.66 | " |
| 1548 | 319 | 49 | 95 | 8.25 | " |
| 1540 | 319 | 46 | 103 | 7.95 | " |
| 1570 | 319 | 49 | 49 | 7.16 | " |
| 1514 | 319 | 50 | 50 | 7.31 | " |
| 1530 | 319 | 50 | 55 | 6.83 | " |
| 1527 | 319 | 50 | 58 | 7.33 | " |
| 1524 | 319 | 50 | 61 | 7.22 | " |
| 1577 | 319 | 50 | 45 | 6.83 | " |
| 1574 | 319 | 50 | 59 | 7.52 | " |
| 1554 | 319 | 50 | 57 | 7.39 | " |
| 1541 | 319 | 50 | 80 | 6.24 | " |
| 1543 | 319 | 50 | 106 | 7.38 | " |
| 1535 | 319 | 50 | 109 | 6.59 | " |
| 1548 | 319 | 50 | 94 | 7.41 | " |
| 1540 | 320 | 47 | 98 | 7.45 | " |
| 1561 | 320 | 49 | 101 | 7.19 | " |
| 1562 | 320 | 49 | 110 | 7.21 | " |
| 1573 | 321 | 52 | 45 | 7.91 | " |
| 1564 | 321 | 52 | 45 | 7.79 | " |
| 1572 | 321 | 52 | 50 | 7.36 | " |
| 1570 | 321 | 52 | 67 | 6.90 | " |
| 1514 | 321 | 53 | 65 | 6.99 | " |
| 1556 | 321 | 77 | 49 | 8.08 | " |
| 1530 | 321 | 53 | 104 | 7.69 | " |
| 1527 | 321 | 53 | 110 | 7.58 | " |
| 1524 | 321 | 53 | 101 | 7.97 | " |
| 1577 | 321 | 53 | 105 | 7.73 | " |
| 1574 | 321 | 53 | 109 | 7.99 | " |
| 1582 | 321 | 77 | 49 | 7.08 | " |
| 1522 | 321 | 77 | 65 | 7.02 | " |
| 1581 | 321 | 76 | 66 | 7.59 | " |
| 1526 | 321 | 77 | 105 | 7.11 | " |
| 1584 | 321 | 77 | 109 | 7.39 | " |
| 1523 | 321 | 77 | 95 | 7.43 | " |
| 1576 | 321 | 77 | 98 | 8.21 | " |
| 1580 | 322 | 77 | 108 | 7.59 | " |
| 1591 | 322 | 77 | 110 | 7.45 | " |
| 1565 | 323 | 79 | 47 | 6.63 | " |
| 1560 | 323 | 79 | 50 | 7.09 | " |
| 1556 | 323 | 79 | 65 | 7.01 | " |
| 1557 | 323 | 79 | 92 | 7.33 | " |
| 1567 | 323 | 79 | 105 | 7.67 | " |
| 1558 | 323 | 79 | 109 | 8.11 | " |
| 1569 | 323 | 79 | 99 | 7.11 | " |
| 1555 | 323 | 79 | 109 | 7.56 | " |
| 1582 | 323 | 79 | 112 | 7.72 | " |
| 1522 | 323 | 79 | 116 | 7.66 | " |
| 1581 | 323 | 78 | 119 | 7.87 | " |
| 1526 | 323 | 79 | 122 | 7.91 | " |
| 1663 | 333 | 2 | 93 | 7.23 | " |
| 1678 | 333 | 3 | 112 | 8.03 | " |
| 1696 | 333 | 3 | 104 | 7.03 | " |
| 1659 | 333 | 2 | 108 | 8.01 | " |
| 1676 | 333 | 2 | 113 | 7.95 | " |
| 1644 | 333 | 2 | 116 | 8.19 | " |
| 1698 | 333 | 2 | 119 | 7.05 | " |
| 1672 | 333 | 2 | 122 | 7.59 | " |
| 1633 | 333 | 7 | 36 | 6.73 | " |
| 1616 | 333 | 7 | 69 | 7.28 | " |
| 1636 | 333 | 8 | 52 | 7.93 | " |
| 1631 | 337 | 12 | 69 | 7.53 | " |
| 1593 | 337 | 57 | 52 | 7.29 | " |
| 1618 | 337 | 57 | 57 | 7.86 | " |
| 1590 | 337 | 57 | 35 | 7.06 | " |
| 1588 | 337 | 57 | 63 | 6.99 | " |
| 1595 | 337 | 57 | 33 | 5.92 | " |
| 1640 | 337 | 57 | 47 | 7.05 | " |
| 1611 | 337 | 57 | 78 | 7.54 | " |
| 1641 | 337 | 57 | 78 | 7.67 | " |
| 1597 | 337 | 57 | 56 | 7.86 | " |
| 1615 | 337 | 57 | 57 | 7.77 | " |
| 1629 | 337 | 57 | 99 | 7.01 | " |
| 1608 | 337 | 57 | 90 | 7.57 | " |
| 1628 | 337 | 57 | 96 | 7.41 | " |
| 1592 | 337 | 57 | 99 | 7.78 | " |
| 1771 | 347 | 52 | 62 | 5.93 | " |
| 1752 | 347 | 50 | 44 | 7.11 | " |
| 1617 | 347 | 48 | 72 | 7.30 | " |
| 1654 | 347 | 49 | 43 | 6.87 | " |
| 1690 | 347 | 48 | 88 | 6.99 | " |
| 1727 | 347 | 48 | 82 | 7.62 | " |
| 1689 | 347 | 48 | 108 | 7.49 | " |
| 1661 | 347 | 48 | 79 | 8.43 | " |
| 1719 | 347 | 48 | 105 | 6.41 | " |
| 1762 | 347 | 48 | 103 | 7.61 | " |
| 1723 | 349 | 52 | 34 | 6.63 | " |
| 1673 | 349 | 51 | 47 | 5.49 | " |
| 1656 | 349 | 52 | 84 | 7.26 | " |
| 1770 | 349 | 51 | 89 | 6.35 | " |
| 1771 | 349 | 54 | 62 | 6.06 | " |
| 1752 | 349 | 52 | 104 | 6.81 | " |
| 1617 | 349 | 50 | 107 | 6.91 | " |
| 1654 | 349 | 51 | 134 | 7.17 | " |
| 1690 | 349 | 50 | 133 | 7.37 | " |
| 1727 | 349 | 50 | 65 | 7.73 | " |
| 1586 | 352 | 25 | 66 | 7.99 | " |
| 1631 | 352 | 27 | 67 | 7.66 | " |
| 1625 | 352 | 26 | 43 | 7.18 | " |
| 1602 | 352 | 26 | 56 | 7.11 | " |
| 1633 | 352 | 26 | 61 | 7.92 | " |
| 1616 | 352 | 26 | 64 | 7.21 | " |
| 1636 | 352 | 27 | 81 | 7.17 | " |
| 1621 | 352 | 24 | 100 | 7.31 | " |
| 1635 | 352 | 26 | 107 | 6.39 | " |
| 1620 | 352 | 24 | 103 | 6.88 | " |
| 1607 | 352 | 23 | 98 | 6.78 | " |
| 1593 | 357 | 77 | 56 | 5.48 | " |
| 1618 | 357 | 77 | 58 | 5.79 | " |
| 1590 | 357 | 77 | 64 | 5.32 | " |
| 1588 | 357 | 77 | 66 | 5.40 | " |
| 1595 | 357 | 77 | 83 | 7.89 | " |
| 1640 | 357 | 77 | 86 | 8.15 | " |
| 1611 | 357 | 77 | 100 | 7.91 | " |
| 1641 | 357 | 77 | 103 | 7.66 | " |
| 1597 | 357 | 77 | 109 | 7.96 | " |
| 1615 | 357 | 77 | 110 | 7.58 | " |
| 1629 | 357 | 77 | 116 | 7.59 | " |
| 1538 | 363 | 45 | 70 | 7.95 | " |
| 1525 | 363 | 44 | 67 | 8.25 | " |
| 1529 | 363 | 42 | 77 | 7.81 | " |
| 1571 | 363 | 45 | 85 | 7.95 | " |
| 1552 | 363 | 45 | 88 | 7.53 | " |
| 1579 | 363 | 45 | 87 | 7.81 | " |
| 1532 | 363 | 45 | 89 | 7.69 | " |
| 1539 | 363 | 45 | 91 | 7.86 | " |
| 1551 | 364 | 46 | 46 | 8.01 | " |
| 1568 | 364 | 46 | 53 | 6.81 | " |
| 1559 | 364 | 46 | 71 | 7.93 | " |
| 1515 | 364 | 46 | 69 | 8.12 | " |
| 1538 | 364 | 46 | 89 | 8.11 | " |
| 1525 | 364 | 45 | 88 | 7.71 | " |
| 1529 | 364 | 43 | 90 | 7.65 | " |
| 1571 | 364 | 46 | 67 | 7.21 | " |
| 1514 | 323 | 54 | 119 | 9.13 | " |
| 1598 | 352 | 25 | 67 | 8.87 | " |
| 1552 | 364 | 46 | 112 | 8.87 | " |
| 1523 | 251 | 7 | 74 | 7.53 | Respler |
| 1580 | 251 | 7 | 66 | 7.81 | " |
| 1534 | 251 | 7 | 61 | 7.37 | " |
| 1560 | 251 | 7 | 84 | 6.75 | " |
| 1556 | 251 | 7 | 93 | 5.83 | " |
| 1591 | 252 | 8 | 80 | 6.24 | " |
| 1544 | 252 | 8 | 88 | 5.13 | " |
| 1581 | 253 | 8 | 79 | 5.69 | " |
| 1526 | 253 | 9 | 78 | 6.40 | " |
| 1584 | 253 | 9 | 87 | 7.05 | " |
| 1576 | 253 | 9 | 79 | 5.33 | " |
| 1554 | 275 | 6 | 109 | 6.03 | " |
| 1541 | 275 | 6 | 96 | 6.67 | " |
| 1540 | 275 | 3 | 59 | 2.72 | " |
| 1561 | 275 | 5 | 59 | 7.33 | " |
| 1564 | 275 | 5 | 54 | 7.26 | " |
| 1565 | 275 | 31 | 59 | 7.61 | " |
| 1514 | 275 | 6 | 57 | 7.05 | " |
| 1557 | 275 | 31 | 62 | 8.51 | " |
| 1436 | 280 | 52 | 68 | 7.48 | " |
| 1611 | 288 | 9 | 62 | 6.45 | " |
| 1641 | 288 | 9 | 68 | 6.49 | " |
| 1597 | 288 | 9 | 74 | 7.79 | " |
| 1615 | 289 | 9 | 77 | 8.07 | " |
| 1629 | 289 | 9 | 82 | 7.51 | " |
| 1608 | 289 | 9 | 87 | 8.22 | " |
| 1628 | 289 | 9 | 85 | 7.51 | " |
| 1592 | 289 | 9 | 90 | 7.68 | " |
| 1595 | 308 | 28 | 76 | 6.79 | " |
| 1601 | 309 | 29 | 107 | 6.66 | " |
| 1630 | 309 | 29 | 101 | 7.96 | " |
| 1619 | 309 | 29 | 80 | 8.39 | " |
| 1634 | 309 | 29 | 74 | 8.15 | " |
| 1593 | 309 | 29 | 68 | 8.39 | " |
| 1618 | 309 | 29 | 64 | 7.69 | " |
| 1524 | 318 | 49 | 63 | 8.64 | " |
| 1577 | 318 | 49 | 64 | 8.79 | " |
| 1582 | 318 | 75 | 89 | 8.41 | " |
| 1535 | 318 | 49 | 105 | 7.05 | " |
| 1548 | 319 | 49 | 110 | 7.36 | " |
| 1657 | 319 | 21 | 92 | 7.49 | " |
| 1752 | 319 | 20 | 59 | 8.80 | " |
| 1654 | 319 | 21 | 56 | 8.11 | " |
| 1690 | 319 | 20 | 56 | 7.51 | " |
| 1727 | 319 | 20 | 46 | 6.88 | " |
| 1661 | 319 | 20 | 50 | 6.76 | " |
| 1567 | 321 | 77 | 69 | 8.59 | " |
| 1558 | 321 | 77 | 75 | 7.59 | " |
| 1569 | 321 | 77 | 82 | 6.36 | " |
| 1555 | 321 | 77 | 84 | 7.62 | " |
| 1522 | 321 | 77 | 93 | 7.53 | " |
| 1538 | 350 | 32 | 80 | 7.95 | " |
| 1525 | 350 | 31 | 78 | 8.43 | " |
| 1552 | 350 | 32 | 73 | 7.59 | " |
| 1579 | 350 | 32 | 68 | 7.34 | " |
| 1532 | 350 | 32 | 70 | 7.66 | " |
| 1539 | 350 | 32 | 66 | 6.97 | " |
| 1565 | 252 | 8 | 79 | 9.43 | " |
| 1595 | 288 | 9 | 61 | 9.59 | " |
| 1601 | 289 | 9 | 100 | 9.09 | " |
| 1522 | 318 | 75 | 88 | 8.83 | " |
| 1436 | 236 | 8 | 76 | 7.64 | MMT |
| 1436 | 237 | 9 | 60 | 7.60 | " |
| 1436 | 265 | 37 | 88 | 7.75 | " |
| 1436 | 278 | 50 | 108 | 7.32 | " |
| 1436 | 341 | 113 | 107 | 7.59 | " |
| 1556 | 245 | 1 | 84 | 7.56 | " |
| 1556 | 248 | 4 | 77 | 5.84 | " |
| 1557 | 247 | 3 | 81 | 7.65 | " |
| 1557 | 248 | 4 | 70 | 6.00 | " |
| 1558 | 247 | 3 | 90 | 6.89 | " |
| 1565 | 245 | 1 | 81 | 2.87 | " |
| 1567 | 247 | 3 | 85 | 7.43 | " |
| 1534 | 278 | 34 | 107 | 6.94 | " |
| 1544 | 278 | 34 | 98 | 7.17 | " |
| 1565 | 278 | 34 | 95 | 7.34 | " |
| 1771 | 298 | 3 | 45 | 5.59 | " |
| 1771 | 299 | 4 | 93 | 6.11 | " |
| 1656 | 298 | 1 | 45 | 3.09 | " |
| 1723 | 298 | 1 | 39 | 6.66 | " |
| 1723 | 299 | 2 | 54 | 7.48 | " |
| 1752 | 298 | 1 | 45 | 6.74 | " |
| 1752 | 299 | 2 | 94 | 7.51 | " |
| 1657 | 299 | 1 | 51 | 7.19 | " |
| 1673 | 298 | 0 | 39 | 5.72 | " |
| 1673 | 299 | 1 | 58 | 6.23 | " |
| 1617 | 299 | 0 | 99 | 7.59 | " |
| 1767 | 300 | 2 | 39 | 6.56 | " |
| 1523 | 304 | 60 | 89 | 7.36 | " |
| 1576 | 304 | 60 | 90 | 7.10 | " |
| 1591 | 304 | 60 | 104 | 6.97 | " |
| 1562 | 301 | 31 | 92 | 7.04 | " |
| 1583 | 300 | 30 | 95 | 7.20 | " |
| 1514 | 336 | 66 | 78 | 7.55 | " |
| 1570 | 336 | 65 | 82 | 7.53 | " |
| 1572 | 335 | 64 | 96 | 4.08 | " |
| 1573 | 335 | 64 | 103 | 7.67 | " |
| 1515 | 334 | 16 | 96 | 7.33 | " |
| 1525 | 335 | 16 | 96 | 7.27 | " |
| 1529 | 334 | 13 | 85 | 7.18 | " |
| 1529 | 335 | 14 | 94 | 7.45 | " |
| 1532 | 336 | 18 | 52 | 6.76 | " |
| 1533 | 317 | 2 | 58 | 6.99 | " |
| 1533 | 332 | 17 | 94 | 7.43 | " |
| 1533 | 334 | 19 | 76 | 7.68 | " |
| 1538 | 334 | 16 | 90 | 7.44 | " |
| 1538 | 335 | 17 | 100 | 7.25 | " |
| 1539 | 333 | 15 | 88 | 7.30 | " |
| 1549 | 332 | 13 | 102 | 7.10 | " |
| 1549 | 333 | 14 | 84 | 7.49 | " |
| 1552 | 336 | 18 | 56 | 7.79 | " |
| 1563 | 334 | 15 | 80 | 3.31 | " |
| 1566 | 332 | 13 | 98 | 7.28 | " |
| 1566 | 334 | 15 | 80 | 2.90 | " |
| 1588 | 309 | 28 | 97 | 7.32 | " |
| 1608 | 323 | 42 | 92 | 7.68 | " |
| 1640 | 309 | 28 | 85 | 7.47 | " |
| 1656 | 335 | 38 | 98 | 7.21 | " |
| 1654 | 319 | 21 | 102 | 7.21 | " |
| 1654 | 335 | 37 | 92 | 7.57 | " |
| 1657 | 300 | 2 | 44 | 6.65 | " |
| 1673 | 300 | 2 | 50 | 5.82 | " |
| 1673 | 335 | 37 | 104 | 5.97 | " |
| 1617 | 319 | 20 | 104 | 7.39 | " |
| 1617 | 335 | 36 | 90 | 7.77 | " |
| 1661 | 319 | 20 | 93 | 7.48 | " |
| 1661 | 334 | 35 | 84 | 7.63 | " |
| 1665 | 319 | 20 | 95 | 7.45 | " |
| 1665 | 335 | 36 | 94 | 7.69 | " |
| 1689 | 319 | 20 | 95 | 7.25 | " |
| 1689 | 334 | 35 | 84 | 7.81 | " |
| 1690 | 319 | 20 | 99 | 6.97 | " |
| 1690 | 335 | 36 | 89 | 7.16 | " |
| 1771 | 300 | 5 | 102 | 6.82 | " |
| 1771 | 319 | 24 | 101 | 6.15 | " |
| 1771 | 335 | 40 | 95 | 5.81 | " |
| 1723 | 300 | 3 | 50 | 6.79 | " |
| 1752 | 300 | 2 | 104 | 7.47 | " |
| 1770 | 335 | 37 | 93 | 5.73 | " |
| 1719 | 318 | 19 | 97 | 7.10 | " |
| 1719 | 319 | 20 | 90 | 7.42 | " |
| 1721 | 319 | 20 | 91 | 7.45 | " |
| 1727 | 319 | 20 | 97 | 7.41 | " |
| 1727 | 320 | 21 | 79 | 7.76 | " |
| 1762 | 318 | 19 | 90 | 7.29 | " |
| 1762 | 335 | 36 | 95 | 7.49 | " |
| 1526 | 340 | 97 | 87 | 6.90 | " |
| 1526 | 341 | 98 | 90 | 7.37 | " |
| 1534 | 340 | 97 | 95 | 7.67 | " |
| 1544 | 340 | 97 | 94 | 7.88 | " |
| 1565 | 340 | 97 | 90 | 7.94 | " |
| 1582 | 340 | 97 | 104 | 6.34 | " |
| 1591 | 340 | 97 | 97 | 2.97 | " |
| 1581 | 340 | 96 | 93 | 6.45 | " |
| 1549 | 346 | 27 | 102 | 7.63 | " |
| 1588 | 340 | 60 | 88 | 7.73 | " |
| 1590 | 340 | 60 | 79 | 7.71 | " |
| 1595 | 340 | 60 | 85 | 7.49 | " |
| 1597 | 341 | 61 | 109 | 7.28 | " |
| 1640 | 340 | 60 | 85 | 7.74 | " |
| 1631 | 336 | 11 | 75 | 7.51 | " |
| 1602 | 336 | 10 | 78 | 5.91 | " |
| 1625 | 336 | 10 | 73 | 7.61 | " |
| 1696 | 346 | 16 | 107 | 7.71 | " |
| 1663 | 346 | 15 | 91 | 7.48 | " |
| 1556 | 353 | 109 | 98 | 7.21 | " |
| 1556 | 354 | 110 | 98 | 7.79 | " |
| 1558 | 355 | 111 | 80 | 7.36 | " |
| 1560 | 353 | 109 | 99 | 7.23 | " |
| 1560 | 354 | 110 | 100 | 7.60 | " |
| 1565 | 353 | 109 | 101 | 7.41 | " |
| 1565 | 354 | 110 | 96 | 7.46 | " |
| 1548 | 354 | 85 | 86 | 7.34 | " |
| 1562 | 353 | 83 | 94 | 7.90 | " |
| 1562 | 354 | 84 | 88 | 7.96 | " |
| 1564 | 353 | 83 | 96 | 7.80 | " |
| 1564 | 354 | 84 | 89 | 7.76 | " |
| 1540 | 354 | 82 | 85 | 7.59 | " |
| 1540 | 355 | 83 | 82 | 7.91 | " |
| 1515 | 350 | 32 | 103 | 6.98 | " |
| 1532 | 350 | 32 | 91 | 7.86 | " |
| 1538 | 350 | 32 | 110 | 5.39 | " |
| 1539 | 350 | 32 | 86 | 7.70 | " |
| 1552 | 350 | 32 | 94 | 7.68 | " |
| 1559 | 350 | 32 | 101 | 7.05 | " |
| 1571 | 350 | 32 | 98 | 7.70 | " |
| 1579 | 350 | 32 | 91 | 7.62 | " |
| 1525 | 350 | 31 | 98 | 6.96 | " |
| 1529 | 350 | 29 | 98 | 7.21 | " |
| 1608 | 353 | 73 | 96 | 7.70 | " |
| 1615 | 353 | 74 | 85 | 7.77 | " |
| 1631 | 353 | 28 | 69 | 6.74 | " |
| 1636 | 353 | 28 | 101 | 7.60 | " |
| 1602 | 353 | 27 | 77 | 6.90 | " |
| 1616 | 353 | 27 | 96 | 7.69 | " |
| 1625 | 353 | 27 | 70 | 6.56 | " |
| 1586 | 353 | 26 | 63 | 7.31 | " |
| 1598 | 353 | 26 | 59 | 7.45 | " |
| 1665 | 348 | 73 | 99 | 7.05 | " |
| 1663 | 363 | 32 | 101 | 7.43 | " |
| 1700 | 363 | 32 | 85 | 7.64 | " |

Appendix B.

Summary of magnitudes below 550 km
'Cal. Day' is the day number of year 2020.
'Ops. Days' (negative) is the time since attaining the 550km operational altitude.

| Starlink Sat # | Cal Day | Ops. Days | Phase Angle | 1000-km Magnitude | Source | Altitude km |
|---|---|---|---|---|---|---|
| 1569 | 231 | -13 | 193 | 7.60 | Mallama | 457 |
| 1555 | 231 | -13 | 193 | 7.69 | " | 458 |
| 1591 | 231 | -13 | 193 | 7.79 | " | 458 |
| 1534 | 231 | -13 | 193 | 8.37 | " | 458 |
| 1571 | 232 | -85 | 265 | 7.80 | " | 380 |
| 1577 | 232 | -36 | 216 | 7.40 | " | 380 |
| 1552 | 232 | -36 | 216 | 7.21 | " | 380 |
| 1574 | 232 | -36 | 216 | 7.42 | " | 380 |
| 1579 | 232 | -85 | 265 | 7.22 | " | 380 |
| 1554 | 232 | -36 | 216 | 7.02 | " | 380 |
| 1532 | 232 | -85 | 265 | 7.23 | " | 380 |
| 1541 | 232 | -36 | 216 | 7.03 | " | 380 |
| 1539 | 232 | -85 | 265 | 7.24 | " | 380 |
| 1543 | 232 | -36 | 216 | 7.04 | " | 380 |
| 1549 | 232 | -86 | 266 | 7.04 | " | 380 |
| 1535 | 232 | -36 | 216 | 7.45 | " | 380 |
| 1548 | 232 | -36 | 216 | 6.86 | " | 380 |
| 1533 | 232 | -1 | 181 | 7.07 | " | 380 |
| 1540 | 232 | -46 | 226 | 6.86 | " | 380 |
| 1563 | 318 | -1 | 181 | 9.36 | " | 549 |
| 1578 | 318 | -1 | 181 | 8.66 | " | 549 |
| 1536 | 318 | -1 | 181 | 8.56 | " | 549 |